\documentstyle[aps,prb]{revtex}

\begin{document}

\input epsf.sty
\twocolumn[\hsize\textwidth\columnwidth\hsize\csname %
@twocolumnfalse\endcsname

\draft

\widetext

\title{Structural instability associated with the tilting of CuO$_{6}$ octahedra 
in La$_{2-x}$Sr$_{x}$CuO$_{4}$}

\author{H. Kimura$^{1,5*}$, K. Hirota$^{1,5}$, C. H. Lee$^{2}$, 
K. Yamada$^{3}$, and G. Shirane$^{4}$}

\address{Department of Physics Tohoku University, Aramaki Aoba,
Sendai 980-8578, Japan}

\address{Physical Science Division, Electrotechnical Laboratory, Umezono, Tsukuba 305-8568, Japan}

\address{Institute for Chemical Research, Kyoto University, Gokasho, Uji 610-0011, Japan}

\address{Department of Physics, Brookhaven National Laboratory, Upton, NY 11973-5000, USA}

\address{CREST, Japan Science and Technology Corporation, Tsukuba 305-0047}

\date{\today}
\maketitle

\begin{abstract}

Comprehensive inelastic neutron-scattering measurements were performed to study 
the soft optical phonons in La$_{2-x}$Sr$_{x}$CuO$_{4}$ at $x=0.10$, 0.12 and 0.18. 
We found at $x=0.18$ that the softening of {\it Z}-point phonon, suggesting incipient 
structural transition from the low-temperature orthorhombic (LTO) to low-temperature 
tetragonal (LTT) phase, {\em breaks} at $T_{\rm c}$, which is consistent with the previous 
report by Lee {\it et al.}\ for the optimally doped $x=0.15$ sample. As for $x=0.10$ and 
0.12, on the other hand, the softening continues even below $T_{\rm c}$. It is 
thus clarified that the {\em breaking} of soft phonon is characteristic of 
La$_{2-x}$Sr$_{x}$CuO$_{4}$ in the optimally and overdoped regions.  
In the course of studying the soft phonons, we discovered that a central peak 
remains above the LTO to high-temperature tetragonal (HTT) phase transition at $T_{\rm s1}$ 
and splits into incommensurate components along the $(1\ 1\ 0)_{\rm HTT}$ direction at higher 
temperatures. This is a common feature for both $x=0.12$ and 0.18 and their temperature 
dependences of the splitting $2\delta$ can be scaled by using a renormalized temperature 
$T/T_{\rm s1}$. In the high temperature limit, $\delta$ saturates around 
$\delta \sim 0.12$~r.l.u., which value is close to the splitting of incommensurate magnetic 
signals. This implies that the incipient lattice modulation starts appearing at very high temperature. 
Details of this modulation and its relations with other properties are, however, not yet clarified. 

\end{abstract}
\vspace*{3mm}


\phantom{.}
]
\narrowtext

\section{Introduction}
\label{intro}

Number of experimental results accumulated for a decade have shown that spin fluctuations, 
appearing as incommensurate peaks around $(\pi,\pi)$, play a significant role in the 
high-$T_{\rm c}$ superconductivity.\cite{Yoshizawa88,Birgeneau89,Cheong91,Yamada95} 
In particular, a recent neutron-scattering work on high-quality 
La$_{2-x}$Sr$_{x}$CuO$_{4}$ (LSCO) single crystals by Yamada {\it et al.}\cite{Yamada98} 
has clearly demonstrated that there exists a linear relationship between the incommensurability 
and $T_{\rm c}$ in a wide hole concentration range. On the other hand, a primary contribution 
of phonon to high-$T_{\rm c}$ superconductivity has not been seriously considered because 
electrical resistivity measurements\cite{Takagi89,Nakamura93} suggest that the electron-phonon 
({\em e-ph}) interaction is not as large as that of BCS superconductors such as A15-type compounds. 

It is known that the structural properties of LSCO is closely related to the tilting pattern of 
CuO$_{6}$ octahedra. Birgeneau {\it et al.}\cite{Birgeneau87} and 
B\"{o}ni {\it et al.}\cite{Boni88} clarified that the tetragonal {\it X}-point phonon 
mode, associated with tilting of CuO$_{6}$ octahedra along the 
[$1\ 1\ 0$] or [$\bar{1}\ 1\ 0$] tetragonal axis, 
goes soft at the high-temperature tetragonal (HTT) to the low-temperature orthorhombic (LTO) 
structural phase transition. Thurston {\it et al}.\cite{Thurston89} studied temperature dependence 
of the $\Gamma$- and {\it Z}-point modes in the LTO phase, which are split from the 
degenerated {\it X}-point mode in the HTT phase. They have shown that the {\it Z}-point mode 
first hardens below the HTT-LTO transition, then softens again with further decreasing 
temperature, indicating incipient structural transition from LTO to the low-temperature 
tetragonal (LTT) phase. The structural phase transitions and soft phonons in this system are 
summarized in Appendix~\ref{appA}.

Recently, Lee {\it et al.}\cite{Lee96} have discovered that the softening of {\it Z}-point 
phonon {\em breaks} at $T_{\rm c}$ in the optimally doped La$_{1.85}$Sr$_{0.15}$CuO$_{4}$ 
sample. This new finding, suggesting a competition between the LTO-LTT transition and 
the appearance of superconductivity, is contrary to the conventional view that the {\em e-ph} 
interaction in LSCO is weak thus not essential. Therefore it is now increasingly 
important to revisit structural properties of LSCO and elucidate their relation to the 
high-$T_{\rm c}$ superconductivity. In the present study, we have carried out comprehensive 
inelastic neutron-scattering measurements of the {\it Z}-point phonon using $x=0.10$, 0.12 
and 0.18 single crystals. Sample characterizations and neutron-scattering measurements are 
described in Sec.~\ref{experi}. To study the phonon softening in detail, we utilized a new technique 
for optimizing resolution function of neutron scattering, which is described in 
Appendix~\ref{appB}. In the course of studying the soft phonons, we discovered 
incommensurate {\em structural} diffuse signals around $(0\ k \pm \delta\ l)_{\rm LTO}$ 
$(k=odd, l=even)$ in the HTT phase for both $x=0.12$ and 0.18. The newly found diffuse 
signals as well as the soft phonons are reported in Sec.~\ref{result}, and discussed in 
Sec.~\ref{discuss}.@

\section{Experimental Details}
\label{experi}

In order to grow a LSCO single crystal which is large, highly crystalline and homogeneous in 
the Sr distribution, Traveling-Solvent-Floating-Zone (TSFZ) method has been utilized and improved 
over past few years.\cite{Lee98} In the present study, we have grown $x=0.10$, 0.12 and 0.18 
single crystals by using the TSFZ method; all the crystals have a volume of about 1~cm$^{3}$ with 
the mosaicness of 0.4$^{\circ}$ full-width at half-maximum (FWHM). To make these crystals 
stoichiometric, i.e., oxygen deficiency free, as-grown samples were annealed under pure oxygen 
gas flow at 900~$^{\circ}$C for 50 hours, then cooled down to 500~$^{\circ}$C with a cooling 
rate of 10~$^{\circ}$C per hour and kept there for 50 hours, and finally cooled down slowly to 
room temperature. 

Bulk magnetic susceptibilities were measured with a SQUID magnetometer. All the samples 
show superconductivity with shielding signal $\geq 100$~\% and the onset of $T_{\rm c}$'s 
were determined to be 28.5~K, 31.5~K and 36.5~K for $x=0.10$, 0.12 and 0.18, respectively. 
The transition width defined as the temperature range between 5~\% and 95~\% of the 
maximum shielding signal is about 3~K for all the samples, which sharpness suggests that doped 
Sr are homogeneously distributed. The {\it c}-axis lengths of all the samples, measured at 
room temperature by using X-ray powder diffractometer, are in good agreement with previous 
report.\cite{Takayama91} The HTT-to-LTO transition temperatures, defined as $T_{\rm s1}$ in 
this paper, were determined to be 240~K and 125~K for $x=0.12$ and 0.18 by neutron diffraction. 

Neutron scattering experiments were performed on the triple-axis spectrometer TOPAN 
installed at the JRR-3M research reactor in the Japan Atomic Energy Research Institute (JAERI). 
The final energy of neutrons was fixed at $E_{\rm f}=14.1$~meV. The $(0\ 0\ 2)$ Bragg  reflection 
of Pyrolytic Graphite (PG) was used in order to monochromatize and analyze neutron beam. 
A typical horizontal collimation is Blank-60$'$-S-60$'$-Blank. PG filter was inserted into 
scattered beam to reduce higher-order contaminations. All the single crystals have twined 
domains in the LTO phase. Therefore, we should consider that all the measurements were 
performed in the superposed $(h\ 0\ l)_{\rm LTO }$ / $(0\ k\ l)_{\rm LTO}$ zone. 
Throughout this paper, 
the reciprocal space is described by using reciprocal lattice unit (r.l.u.) in the LTO ({\it Bmab}) 
coordinate system. A crystal was put into an Aluminum container filled with He gas then attached 
to the cold finger of a closed-cycle $^{4}$He cryostat. Temperature was measured by using Si-diode 
thermometers attached to the cold finger and the bottom part of the Al container. 

\section{Results}
\label{result}

\subsection{Soft phonons}
\label{soft}

The {\it Z}-point phonons on $x=0.10$, 0.12 and 0.18 around ${\bf Q}_{0}=(3\ 0\ 2)$ were 
studied as a function of temperature. Constant-$Q$ scans were carried out at 
${\bf Q'}={\bf Q}_{0}+{\bf q}$ where the instrumental resolution matches most effectively 
with the  dispersion of phonon ({\em focusing}). The procedure we employed to determine the most 
effective focusing point ${\bf Q'}$ is described in Appendix~\ref{appB}. The phonon spectra for 
$x=0.10$ and 0.12 were measured at ${\bf Q'}=(3.09\ 0\ 1.88)$, while for 
$x=0.18$ the scan was carried out at ${\bf Q'}=(3.10\ 0\ 1.96)$. Well-defined phonon 
spectra were obtained for all the samples, indicating that the resolution ellipsoid at 
${\bf Q'}$ sufficiently focuses with the slope of dispersion. To quantitatively analyze the phonon 
spectra, data were fitted with the following scattering function $S(Q,\omega)$ convoluted with a 
proper instrumental resolution:
\begin{eqnarray*}
S(Q,\omega) &=& \frac{\omega}{1 - \exp(-\omega/k_{B}T)}\;
\Bigl\{\frac{1}{\pi}\;\frac{A}{\Gamma_{\rm ph}/2}\;
\frac{1}{1 + \bigl(\frac{\omega - \omega_{\rm ph}}%
{\Gamma_{\rm ph}/2}\bigr)^2}\Bigr\},\\
A &=& \chi(Q) = \int_{-\infty}^{+\infty}\;\frac{\chi(Q,\omega)}{\omega}\;
d\omega,\\
\end{eqnarray*}
where $\omega_{\rm ph}$ and $\Gamma_{\rm ph}$ are the phonon energy and the intrinsic 
line-width at ${\bf Q'}$, respectively. Fitting thus carried out quite well reproduce all the observed 
data.

\begin{figure}
\centerline{\epsfxsize=3in\epsfbox{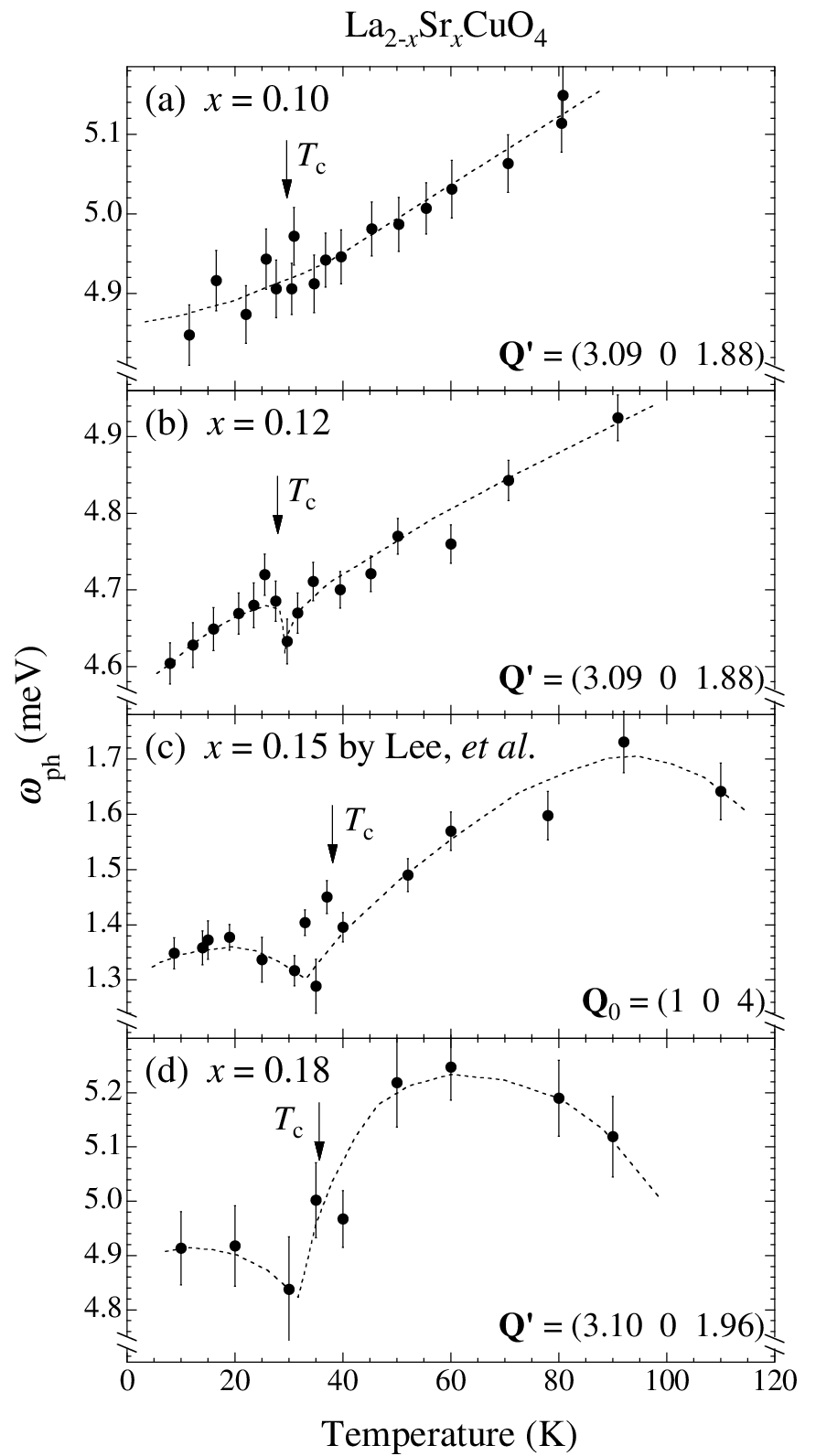}}
\vspace*{3mm}
\caption{Temperature dependence of the soft phonon energy $\omega_{\rm ph}$ 
  for $x=0.10$, 0.12, 0.15, 0.18. Figures (a) and (b) show the data for $x=0.10$ and 0.12, 
  respectively, which were measured at ${\bf Q'}=(3.09\ 0\ 1.88)$. Figure (c) is the data for 
  $x=0.15$ at {\it Z}-point; $(1\ 0\ 4)$, which was taken by 
  Lee {\it et al}.\protect\cite{Lee96}. 
  The data for $x=0.18$ at ${\bf Q'}=(3.10\ 0\ 1.96)$ is shown  in figure (d). 
  Dashed lines in all the figures are guides to the eye.}
\label{fig1}
\end{figure}

Temperature dependences of the soft phonon energy $\omega_{\rm ph}$ for $x=0.10$, 0.12 
and 0.18 are shown in Fig.~\ref{fig1}. All the data exhibit the softening of phonons with 
decreasing temperature toward $T_{\rm c}$, which confirms that the phonons at ${\bf Q'}$ 
indeed reflect the behaviors of {\it Z}-point phonons. As a reference, the {\it Z}-point 
(${\bf Q}_{0}=(1\ 0\ 4)$) phonon for $x=0.15$ reported by Lee {\it et al.}\cite{Lee96} is 
also shown in Fig.~\ref{fig1}(c). 
For $x=0.18$, the softening of phonon {\em breaks} 
with the appearance of superconductivity as seen in $x=0.15$. In contrast, the softening 
continues even below $T_{\rm c}$ for $x=0.10$ and 0.12. These results indicate that 
the structural instability for the LTT phase persists even in the superconducting phase for 
$x=0.10$ and 0.12. 

\begin{figure}
\centerline{\epsfxsize=3in\epsfbox{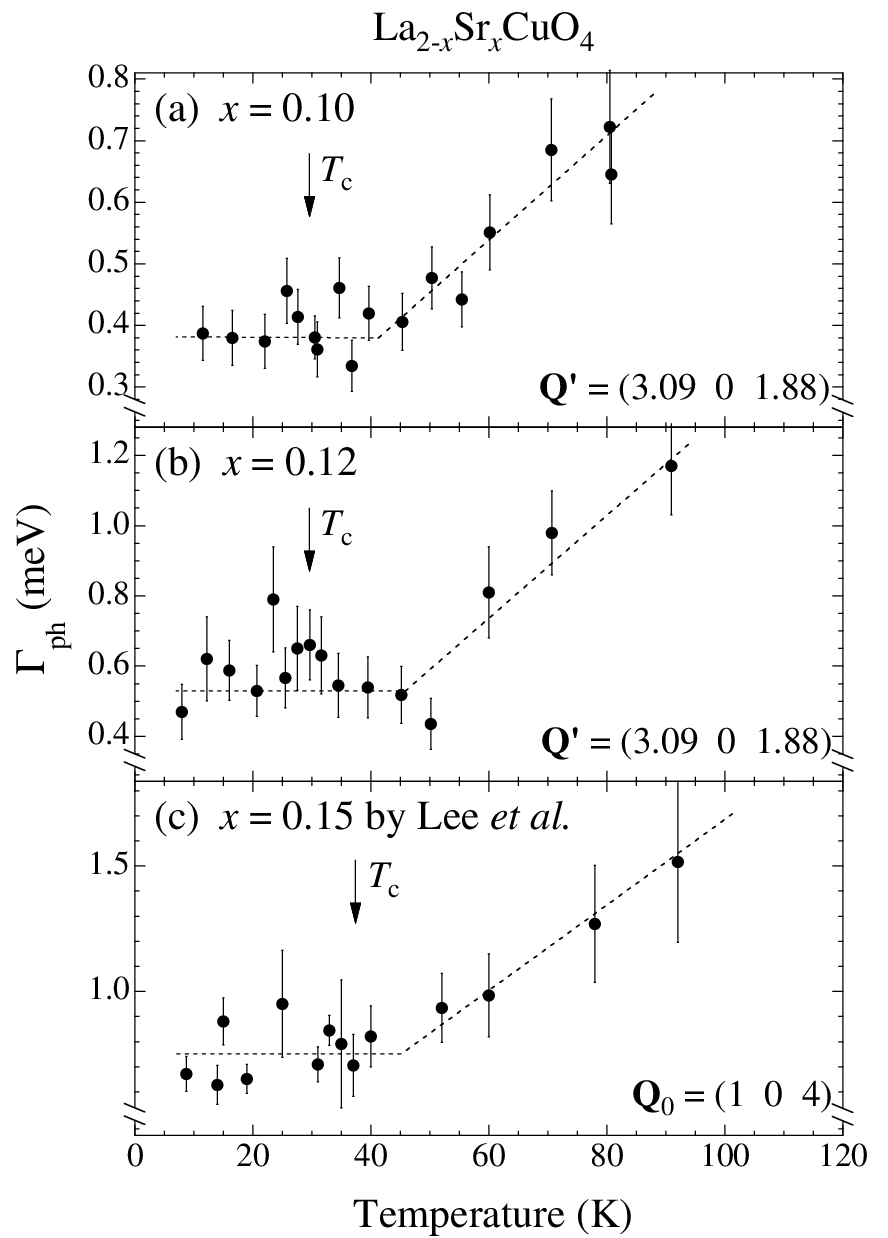}}
\vspace*{3mm}
\caption{Line-width $\Gamma_{\rm ph}$ of the soft phonon for $x=0.10$, 0.12 at 
  ${\bf Q'}=(3.09\ 0\ 1.88)$, and 0.15 at {\it Z}-point are shown as a 
  function of temperature. Figures (a) and (b) correspond to the data for $x=0.10$ and 
  0.12, respectively. The data for $x=0.15$; (c) was measured by 
  Lee {\it et al}.\protect\cite{Lee96}. Dashed lines are guides to the eye.}
\label{fig2}
\end{figure}

The intrinsic line-widths of the soft phonons $\Gamma_{\rm ph}$ for $x=0.10$, 0.12 
and 0.15 are shown as a function of temperature in Figs.~\ref{fig2}(a)-(c). The data for 
$x=0.15$ is also after Lee {\it et al}.\cite{Lee96}. Down to 40 K, the line-widths for all the 
samples decrease with decreasing temperature, corresponding to increase of phonon life-time. 
However, this narrowing suddenly stops, i. e. {\em breaks}, around 40~K ($\equiv T_{d}$) 
and the line-widths become constant at lower temperatures. On the other hand, in 
La$_{2}$CuO$_{4}$ studied by Lee {\it et al}.\cite{Lee96}, the narrowing continues down to 
the lowest temperature, indicating that the {\em break} of the line-width narrowing is a 
characteristic phenomenon of 
\linebreak
\noindent
the Sr-doped samples. As for $x=0.18$, it is difficult to estimate 
the accurate value of $\Gamma_{\rm ph}$ because the two phonons, i.e., $\Gamma$ and 
{\it Z}-point modes, were so close that the phonon spectra were superposed with each other. 

\subsection{Incommensurate diffuse peak in the HTT phase}
\label{incomme}

Upon cooling, the $(0\ k\ l)$  $(k=odd, l=even)$ superlattice reflections emerge at the HTT-LTO 
transition temperature $T_{\rm s1}$. As briefly described in Sec.~\ref{experi}, $T_{\rm s1}$ 
values of $x=0.12$ and $x=0.18$ were determined to be 240~K and 125~K from 
the temperature dependence of the 
\linebreak
\noindent
$(0\ 3\ 2)$ peak intensity. Note that we inserted a PG filter 
between sample and analyzer as well as between monochromator and sample, which almost 
completely eliminates higher-order contaminations. In the course of studying the soft-mode 
phonons and superlattice reflections, we found that there remains a weak diffuse peak centered at 
$(0\ 3\ 2)$ and $\omega=0$ even above $T_{\rm s1}$. Since this peak is associated with 
the softening of {\it X}-point phonon and diverges at $T_{\rm s1}$, we have noticed that 
this is a so-called ``central peak'' as studied in SrTiO$_{3}$\cite{Shapiro72,Shirane93} and many 
other systems.

\begin{figure}
\centerline{\epsfxsize=3in\epsfbox{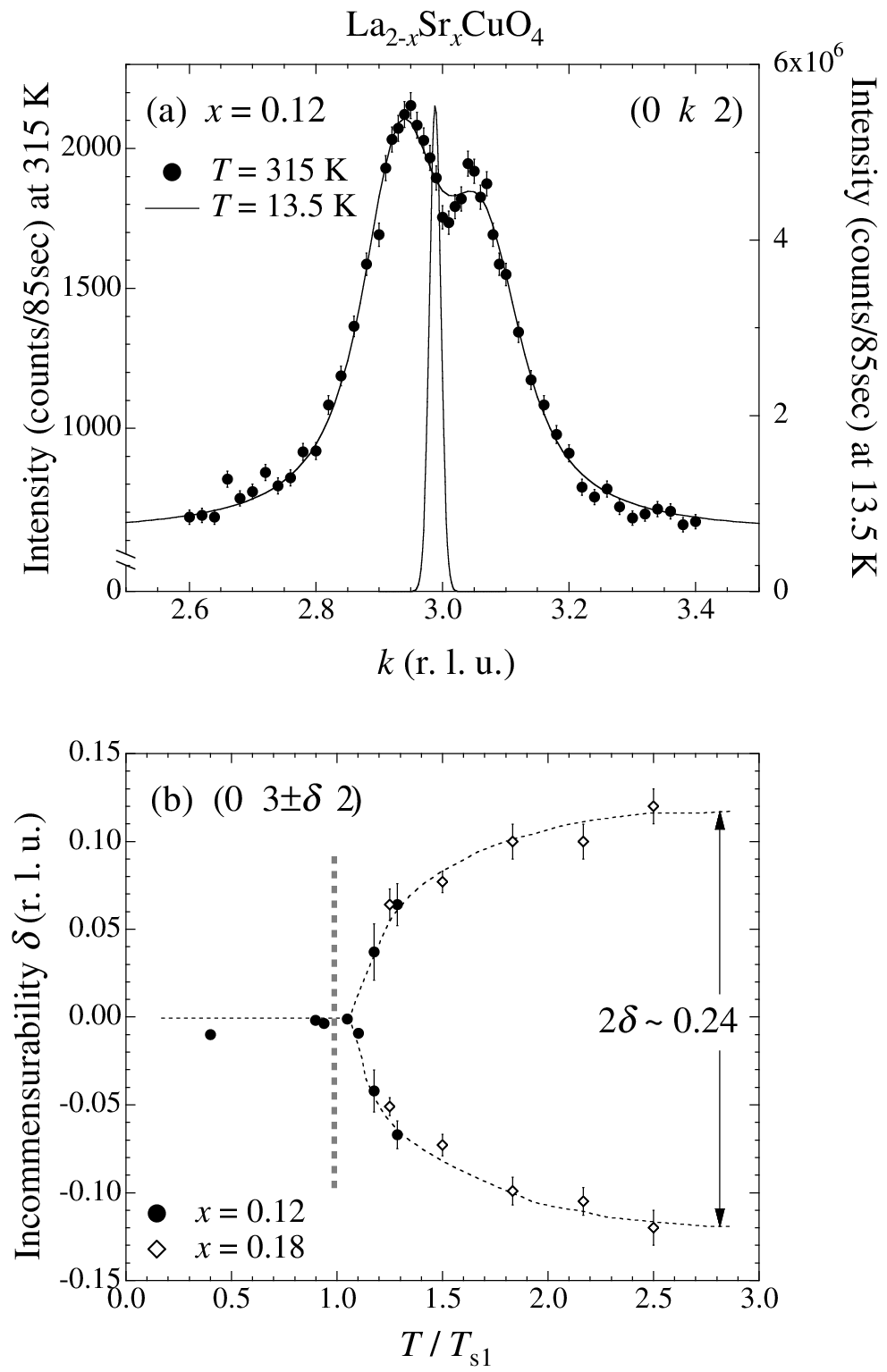}}
\vspace*{3mm}
\caption{(a) Incommensurate diffuse peak spectrum along $(0\ k\ 2)$ for $x=0.12$ at 315~K. 
  The left vertical axis denotes the intensity of the diffuse peak. The sharp single peak with the solid 
   line shows the $(0\ 3\ 2)$ superlattice reflection at 13.5~K. The intensity of the sharp 
   superlattice peak is shown in the right vertical axis. (b) Temperature dependence of the 
   incommensurability $\delta$ for $x=0.12$ and 0.18, which correspond to the peak position of 
   the  incommensurate diffuse peak. The bottom axis is scaled by $T_{\rm s1}$ for $x=0.12$, 
   0.18. Dashed lines are guides to the eye.}
\label{fig3}
\end{figure}

At higher temperature, we found that this central peak starts splitting into two incommensurate 
components at $(0\ 3 \pm\delta\ 2)$ or more generally $(0\ k\pm\delta\ l)$ $(k=odd, l=even)$. 
This phenomenon was first reported by Shirane {\it et al}.\ in their preliminary experiment 
for $x=0.15$.\cite{Shirane**} As shown in Fig.~\ref{fig3}(a), the incommensurate ``central'' 
peaks are clearly seen in $x=0.12$ at 315~K, temperature much higher than $T_{\rm s1}$. 
Similar incommensurate peaks were also observed for $x=0.18$. The peak profiles for both 
samples were well fitted with a double Lorentzian. The sharp peak in Fig.~\ref{fig3}(a) shows the 
$(0\ 3\ 2)$ superlattice peak 
\linebreak
\noindent 
at $T=13.5$~K, of which the line-width corresponds to 
the instrumental resolution at $(0\ 3\ 2)$. By fitting with a double Lorentzian convoluted 
with the resolution, the intrinsic line-widths of central peaks were estimated to be 
$\sim 0.06$~\AA$^{-1}$ for both the samples and for all the temperature range above 
$T_{\rm{s1}}$, indicating short range in-plane correlations ($\xi \sim 17$~\AA). 

The temperature dependence of the splitting $\delta$ for $x=0.12$ and 0.18 were also measured. 
As shown in Fig.~\ref{fig3}(b), the $\delta$ value of $x=0.18$ (open diamonds) increases with 
increasing temperature and is saturated around $\delta \sim 0.12$ (r.l.u.). Upon cooling, on the other 
hand, $\delta$ approaches zero at $T_{\rm s1}$, suggesting that the central peak indeed starts 
splitting just at $T_{\rm s1}$. It is remarkable that the $\delta$ values for both $x=0.12$ 
(closed circles) and 0.18 can be scaled using each $T_{\rm s1}$ value. Note that the saturated 
value $\delta \sim 0.12$ is very close to the incommensurability $\epsilon$ of elastic and 
inelastic magnetic scattering peaks seen in La$_{2-x}$Sr$_{x}$CuO$_{4}$ at 
$(\pi(1 \pm 2\epsilon), \pi)$ and $(\pi, \pi(1 \pm 2\epsilon))$. In the $x=0.25$ sample, 
which has no HTT to LTO structural phase transition, the central peak was also searched around 
$(0\ 3\ 2)$ at room temperature. However, no signature was found.

The incommensurate central peak was also measured around $(0\ 1\ 4)$ at 315~K for $x=0.12$. 
The intrinsic line-width is  almost same as that of $(0\ 3\ 2)$. The width of the central peak along 
the $l$ direction, which corresponds to the out-of-plane correlation, was measured at $(0\ 1\ 4)$. 
The peak-width along {\it l} is almost identical to that along {\it k}, indicating short-range 
isotropic correlations for the in-plane and out-of-plane directions. The intensity ratio of the 
$(0\ 3\ 2)$ and $(0\ 1\ 4)$ central peaks is 2.9, which is close to the ratio of $Q^{2}$ 
values for these two peaks, 2.7, indicating that the central peak originates from atomic 
displacement. The intensity ratio between the central peak and the fundamental Bragg peak is about 
10$^{-3}$ in the present neutron-scattering measurement. We found a similar diffuse peak in X-ray 
diffraction measurements. However, this diffuse peak intensity in X-ray diffraction is 10$^{-5}$ 
times weaker than that of the fundamental reflection. This difference between neutron and X-ray 
measurements suggests that the central peaks are mainly contributed from the displacement of 
oxygen atoms.

\section{Discussion}
\label{discuss}

\subsection{Soft phonons}
\label{soft_d}

The present study, combined with the work by Lee {\it et al.}~\cite{Lee96}, has established that the 
softening of {\it Z}-point phonon suggesting incipient LTO-LTT transition {\em breaks} at 
$T_{\rm c}$ for optimally and overdoped LSCO and that the softening persists even below 
$T_{\rm c}$ in the underdoped region. This {\em breaking} found in optimally and overly doped 
LSCO is consistent with previous results obtained by other 
techniques\cite{Nohara93,Suzuki94,Buchner91,Buchner93} suggesting competition between 
the structural phase transition and the superconductivity in LSCO and related compounds. 
Such a competition is also seen in other systems such as the A15 
superconductors.\cite{Kataoka86,Toyota88} On the contrary, it appears that the behaviors 
of soft phonons in the underdoped region are qualitatively different from those in the overdoped 
region. Note also that the persistence of phonon softening below $T_{\rm c}$ is not a 
characteristic feature for the 1/8-doping. The change in behaviors of phonon softening 
across the optimum hole concentration might be related to the incommensurate elastic 
magnetic peaks in the underdoped region\cite{Kimura99,Matsushita99} and/or the energy-gap in 
the spin fluctuations near the optimally doped region.\cite{Lee99} 
It is intriguing to relate the appearance of elastic magnetic signals to the LTO-to-LTT 
instability persisting below $T_{\rm c}$ in underdoped LSCO. It is clear, however, that 
even more systematic investigation is required to elucidate the origin of soft-phonon 
breaking and its influence to other properties. 

Another remarkable feature is that the narrowing of soft-phonon line-width {\em breaks} below 
$T_{\rm c}$, which is confirmed in underdoped and optimally doped LSCO. This phenomenon is 
contrary to that observed in BCS superconductors, \cite{Axe73,Shapiro75} where {\it e-ph} 
scattering disappears because of the BCS gap opening. It is thus suggested that the intrinsic phonon 
line-width is not owing to {\it e-ph} interactions, though the origin of the breaking of soft-phonon 
narrowing is still not clear. 

\subsection{Incommensurate diffuse peaks}
\label{incomme_d}

Residual week diffuse peak observed above $T_{\rm s1}$ at the LTO superlattice position is very 
similar to the central peak associated with {\it R}-point phonon softening in 
SrTiO$_{3}$.\cite{Shapiro72,Shirane93} Interestingly, the central peak in LSCO splits into 
incommensurate peaks and the incommensurability $\delta$ saturates around 
$\delta \sim 0.12$~r.l.u., which value is close to the splitting of incommensurate magnetic 
signals. This implies that the incipient lattice modulation starts appearing at very high temperature. 
It has been reported that the averaged (long-ranged) structure is different from local 
(short-ranged) structures in the LSCO cuprates.\cite{Bianconi96,Billinge94,Billinge96} 
Quantitative comparison between the experimental results and structural models including local 
distortions is required. Further neutron scattering along this line is planned in the near future.

\section*{Acknowledgments}

We thank Y. Endoh, N. Toyota, R. J. Birgeneau, T. Imai, and H. Fukuyama, for several 
stimulating discussions. We also acknowledge M. Onodera and K. Nemoto for their 
technical supports on neutron scattering experiments. This work was supported in part by a 
Grant-In-Aid for Scientific Research from the Japanese Ministry of Education, Science, Sports 
and Culture, and by a Grant for the Promotion of Science from the Science and Technology Agency 
and also supported by CREST and the US-Japan cooperative research program on Neutron 
Scattering. Work at Brookhaven National Laboratory was carried out under contact 
No. DE-AC02-98CH10886, Division of Material Science, U. S. Department of Energy.

\pagebreak
\appendix


\section{Structural phase transitions and soft phonons in LSCO}
\label{appA}

\begin{figure}
\centerline{\epsfxsize=3in\epsfbox{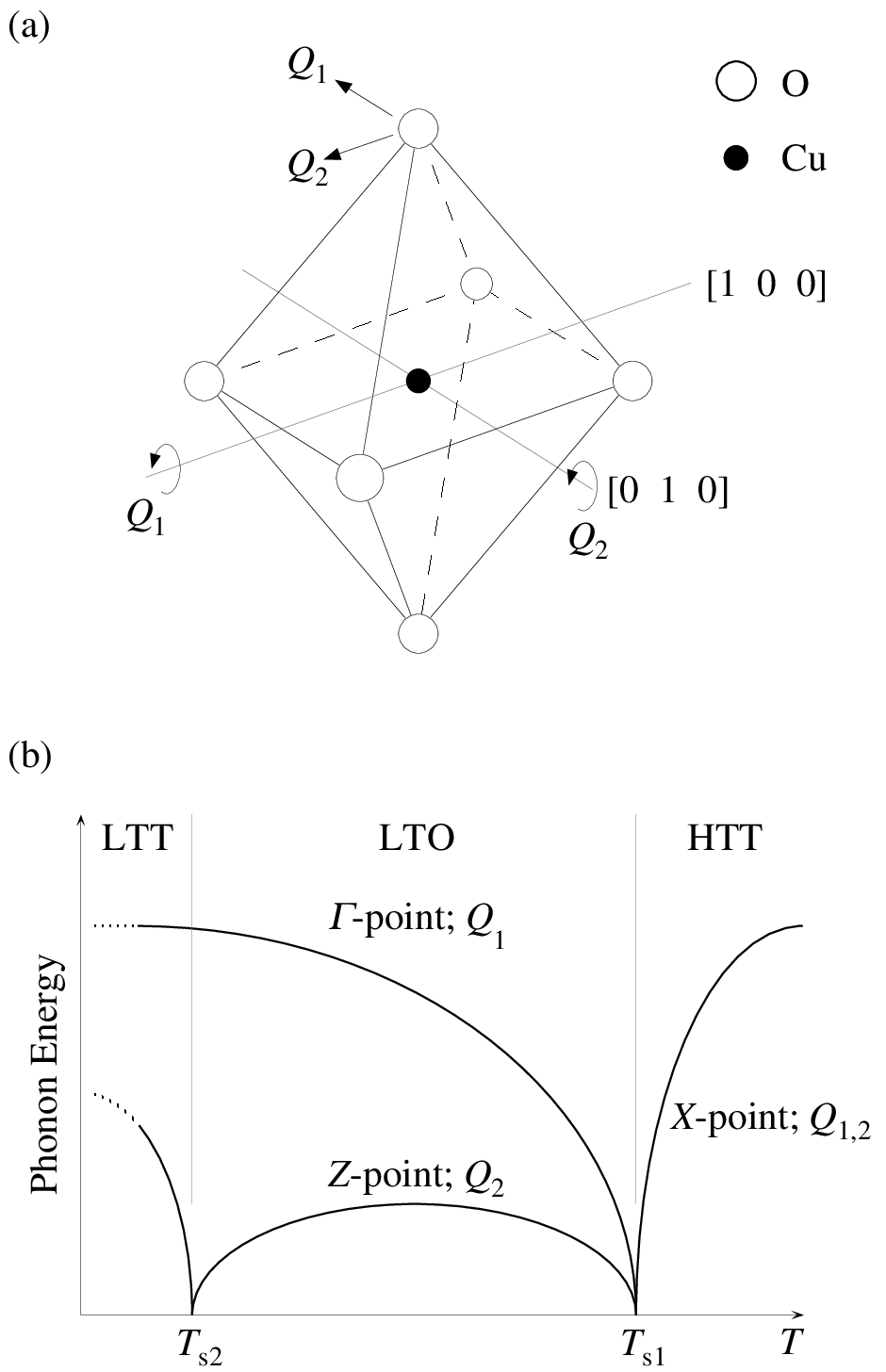}}
\vspace*{3mm}
\caption{(a); Rotation patterns of the CuO$_{6}$ octahedron in LSCO. $Q_{1}$ and $Q_{2}$ 
  are defined as the rotations of $[1\ 0\ 0]_{\rm LTO}$ and $[0\ 1\ 0]_{\rm LTO}$ axes, 
  which correspond to amplitude of soft phonons. (b); Schematic drawing of the temperature 
  dependence of phonon frequencies corresponding to {\it X}-, $\Gamma$-, and 
  {\it Z}-point in the HTT, LTO, and LTT phases.}
\label{fig4}
\end{figure}

As shown in Fig.~\ref{fig4}(a), all the structural phases of LSCO can be characterized by 
two order parameters $Q_{1}$ and $Q_{2}$, which correspond to the rotation of the 
CuO$_{6}$ octahedron along [$1\ 0\ 0$] and [$0\ 1\ 0$] in {\it Bmab} notation. In the 
high-temperature tetragonal (HTT) phase, no static tilting 
exists and the structure has a 
tetragonal symmetry with the space group of {\it I4/mmm}; $|Q_{1}|=|Q_{2}|=0$. 
With decreasing temperature, coherent tilting which rotation axis is parallel to [$1\ 0\ 0$] 
develops below $T_{\rm s1}$ and it leads to the low-temperature orthorhombic phase (LTO) 
with the space group {\it Bmab}; $|Q_{1}|\neq 0, |Q_{2}|=0$. $T_{\rm s1}$ changes from 
520~K for $x=0.00$ to 0~K for $x>0.21$. In some doped LaCuO$_{4}$ such as 
La$_{2-x}$Ba$_{x}$CuO$_{4}$, with further lowering temperature, a low-temperature 
tetragonal (LTT) phase appears with the space group 
{\it P4}$_{\it 2}${\it /ncm}; $|Q_{1}|=|Q_{2}|\neq 0$. 
An intermediate phase 
\linebreak
\noindent 
between LTO and LTT, which is written in the {\it Pccn} orthorhombic 
symmetry, is also known;  $|Q_{1,2}|\neq 0, |Q_{1}|\neq |Q_{2}|$. 

The HTT-to-LTO structural phase transition in LSCO is driven by softening of the zone-boundary 
optical phonon at {\it X}-point ${\bf q} = \frac{1}{2}(1\ \pm 1\ 0)$\cite{Birgeneau87,Boni88}. 
The temperature dependence of the soft phonon energy is shown schematically in 
Fig.~\ref{fig4}(b), which is based on the Landau free energy expanding to eighth order for 
\{$Q_{1}$, $Q_{2}$\}\cite{Wu90,Wu91}. At $T_{\rm s1}$ the doubly degenerate {\it X}-point 
phonons freeze, resulting in the HTT-to-LTO phase transition. In the LTO phase, due to the 
crystal symmetry, the degenerate phonon branches split into $\Gamma$- and {\it Z}-point 
phonon of which amplitudes correspond to $Q_{1}$ and $Q_{2}$, respectively. 
The $\Gamma$-point phonon hardens with decreasing temperature suggesting the stability of 
coherent $|Q_{1}|$ tilting, while the {\it Z}-point phonon softens at lower temperature, indicating 
the {\em instability} toward the LTT phase. If the {\it Z}-point phonon completely freezes, 
a further transition from the LTO to {\it Pccn} or LTT phase occurs. As for LSCO, no 
{\it Pccn} or LTT phase exists but the softening of the {\it Z}-point phonon is 
observed\cite{Thurston89,Lee96}, indicating incipient transition to these phases. 
Note that since the present study cannot distinguish a qualitative difference between the {\it Pccn} 
and LTT, we define a structural phase below LTO as the LTT phase in this paper for convenience. 


\section{Experimental techniques in phonon measurements}
\label{appB}

\begin{figure}
\centerline{\epsfxsize=3in\epsfbox{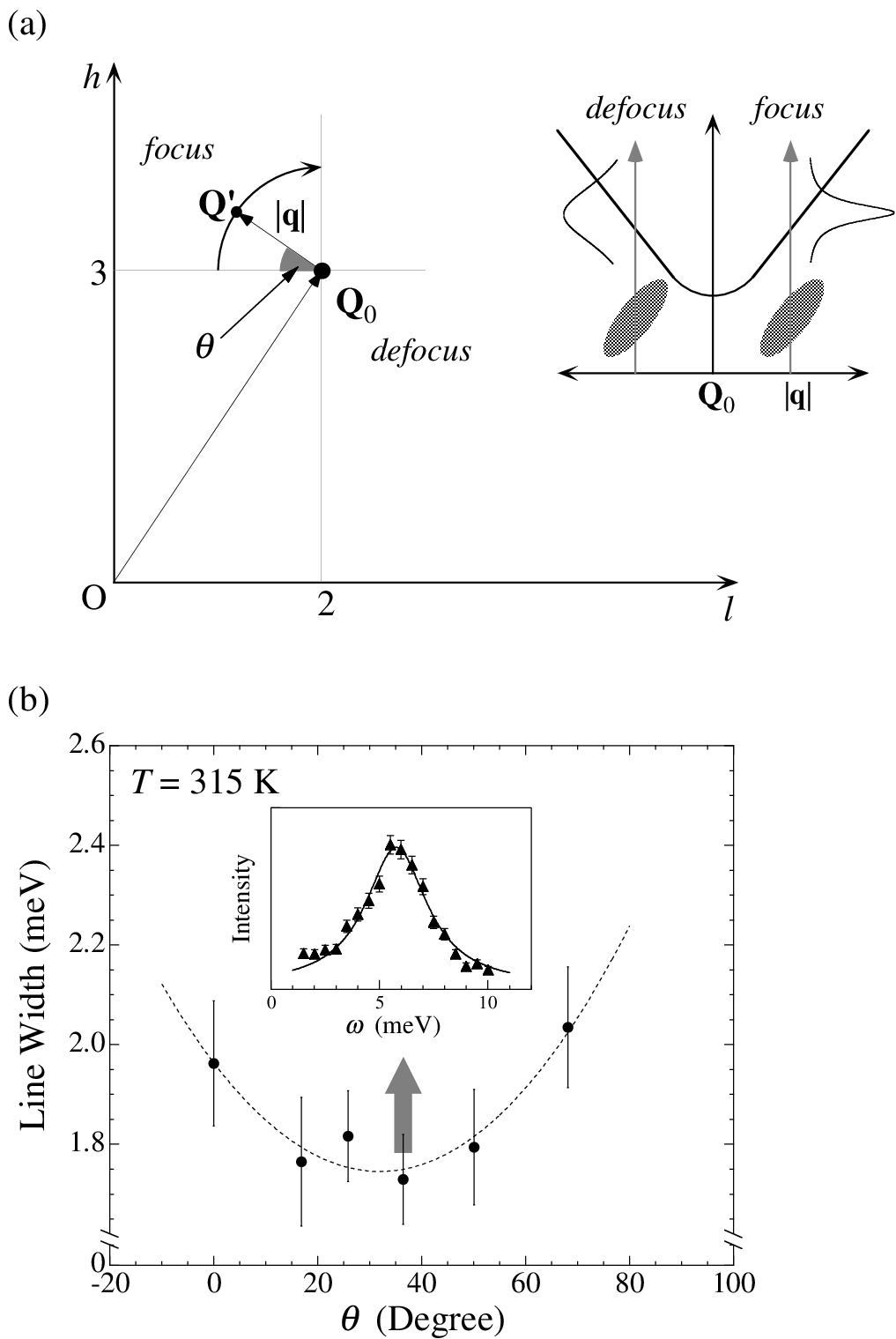}}
\vspace*{3mm}
\caption{(a); The reciprocal lattice space of $(h\ 0\ l)$ for LSCO. 
  ${\bf Q}_{0}(=(3\ 0\ 2))$ is the $Z$-point in the LTO phase. Constant-$Q$ scans at 
  ${\bf Q'}={\bf Q}_{0}+{\bf q}$ were carried out on the circular arc shown in the figure. 
  $\theta$ is defined as the angle between $[0\ 0\ 1]$ axis and $q$. The inset of the figure (a) 
  shows an observed spectrum under the {\em focusing} or {\em defocusing} condition. 
  The schematic bold curve shows a dispersion curve of a soft phonon and grey ellipsoids show 
  the instrumental resolution at ${\bf Q}={\bf Q}_{0}\pm {\bf q}$. (b); Observed line-widths 
  of the phonon spectrums at ${\bf Q'}$ as a function of $\theta$. The inset shows the phonon 
  spectrum at $\theta=37^{\circ}$ where the line-width has a minimum value.}
\label{fig5}
\end{figure}

For the present study, it is essential to precisely measure the frequency and intrinsic 
width of phonon. As 
shown in the inset of Fig.~\ref{fig5}(a), the instrumental resolution 
of neutron triple-axis spectrometer has a slope in the $Q-\omega$ space. Therefore, there 
exist {\em focusing} and {\em defocusing} side for the phonon dispersion. Instead of performing a 
constant-$Q$ scan at the zone-boundary $X$-point ${\bf Q}_{0}$, we looked for the most 
effective focusing point in the vicinity of $Q_{0}$. We made constant-$Q$ scans at several 
different position ${\bf Q'}$ on the arc ${\bf Q'}={\bf Q}_{0}+{\bf q}$ as shown in 
Fig.~\ref{fig5}(a), making a natural assumption that the phonon dispersion is isotropic 
near ${\bf Q_{0}}$. Figure~\ref{fig5}(b) shows the $\theta$ dependence of the line-width 
of the phonon spectrum for $x=0.12$ at $T=315$~K, where $\theta$ is defined as shown in 
Fig.~\ref{fig5}(a). It is clearly seen that the line-width has a minimum at 
$\theta \sim 37^{\circ}$. We have thus chosen ${\bf Q'}=(3.09\ 0\ 1.88)$ for $x=0.10$ and 
0.12. The well-defined phonon spectrum for $x=0.12$ obtained at ${\bf Q'}$ is shown in the inset 
of Fig.~\ref{fig5}(b). As for $x=0.18$, ${\bf Q'}$ was determined to be $(3.10\ 0\ 1.96)$ by 
using the same procedure.

Note that this method is based upon the assumption that, in the formula of phonon dispersion 
$\omega_{\rm ph}^{2}(q)=\omega^{2}(Q_{0})+Aq^{2}$, the coefficient $A$ is weakly 
temperature dependent. This assumption was justified by the work done by Birgeneau 
{\it et al.}\cite{Birgeneau87}.

\footnotetext[1]{\noindent Present address:  Research Institute for Scientific Measurements, 
Tohoku University, Aoba-ku, Sendai 980-8577, Japan.}


\newpage

\begin{references}

\bibitem{Yoshizawa88} H. Yoshizawa, S. Mitsuda, H. Kitazawa and K. Katsumata: 
J. Phys.\ Soc.\ Japan {\bf 57} (1988) 3686.

\bibitem{Birgeneau89} R. J. Birgeneau, Y. Endoh, Y. Hidaka, K. Kakurai, M. A. Kastner, 
B. Keimer, T. Murakami, G. Shirane, T. R. Thurston and K. Yamada: 
Phys.\ Rev.\ B {\bf 39} (1989) 2868.

\bibitem{Cheong91} S.-W. Cheong, G. Aeppli, T. E. Mason, H. Mook, S. M. Hayden, 
P. C. Canfield, Z. Fisk, K. N. Claussen and J. L. Martinez: 
Phys.\ Rev.\ Lett. {\bf 67} (1991) 1791.

\bibitem{Yamada95} K. Yamada, S. Wakimoto, G. Shirane, C.-H. Lee, M. A. Kastner, 
S. Hosoya, M. Greven, Y. Endoh and R. J. Birgeneau: 
Phys.\ Rev.\ Lett. {\bf 75} (1995) 1626.

\bibitem{Yamada98}K. Yamada, C. H. Lee K. Kurahashi, J. Wada, S. Wakimoto, S. Ueki, H. Kimura, 
Y. Endoh, S. Hosoya, G. Shirane, R. J. Birgeneau, M. Greven, M. A. Kastner and Y. J. Kim: 
Phys.\ Rev.\ B {\bf 57} (1998) 6165.

\bibitem{Takagi89}H. Takagi, T. Ido, S. Ishibashi, M. Uota, S. Uchida and Y. Tokura: 
Phys.\ Rev.\ B {\bf 40} (1989) 2254.

\bibitem{Nakamura93}Y. Nakamura and S. Uchida: 
Phys.\ Rev.\ B {\bf 47} (1993) 8369.

\bibitem{Birgeneau87}R. J. Birgeneau, C. Y. Chen, D. R. Gabbe, H. P. Jenssen, M. A. Kastner, 
C. J. Peters, P. J. Picone, T. Thio, T. R. Thurston, H. L. Tuller, J. D. Axe, P. B\"{o}ni and G. Shirane: 
Phys.\ Rev.\ Lett. {\bf 59} (1987) 1329.

\bibitem{Boni88}P. B\"{o}ni, J. D. Axe, G. Shirane, R. J. Birgeneau, D. R. Gabbe, H. P. Jenssen, 
M. A. Kastner, C. J. Peters, P. J. Picone and T. R. Thurston: 
Phys.\ Rev.\ B {\bf 38} (1988) 185.

\bibitem{Thurston89}T. R. Thurston, R. J. Birgeneau, D. R. Gabbe, H. P. Jenssen, M. A. Kastner, 
P. J. Picone, N. W. Preyer, J. D. Axe, P. B\"{o}ni, G. Shirane, M. Sato, K. Fukuda and S. Shamoto: 
Phys.\ Rev.\ B {\bf 39} (1989) 4327.

\bibitem{Lee96}C. H. Lee K. Yamada, M. Arai, S. Wakimoto, S. Hosoya and Y. Endoh: 
Physica C {\bf 257} (1996) 264.

\bibitem{Lee98}C. H. Lee N. Kaneko, S. Hosoya, K. Kurahashi, S. Wakimoto, K. Yamada and Y. Endoh: 
Supercond.\ Sci.\ Technol. {\bf 11} (1998) 891.

\bibitem{Takayama91}E. Takayama-Muromachi and D. E. Rice: 
Physica C {\bf 177} (1991) 195.

\bibitem{Shapiro72}S. M. Shapiro, J. D. Axe, G. Shirane and T. Riste: 
Phys.\ Rev.\ B {\bf 6} (1972) 4332.  

\bibitem{Shirane93}G. Shirane, R. A. Cowley, M. Matsuda and S. M. Shapiro: 
Phys.\ Rev.\ B {\bf 48} (1993) 15595.

\bibitem{Shirane**}G. Shirane {\it et al.}Unpublished. 

\bibitem{Nohara93}M. Nohara, T. Suzuki, Y. Maeno and T. Fujita: 
Phys.\ Rev.\ Lett. {\bf 70} (1993) 3447. 

\bibitem{Suzuki94}T. Suzuki, M. Nohara, Y. Maeno, T. Fujita, T. Tanaka and H. Kojima: 
J. Supercond. {\bf 7} (1994) 419.

\bibitem{Buchner91}B. B\"{u}chner, M. Braden, M. Cramm, W. Schlabitz, W. Schnelle, O. Hoffels, 
W. Braunisch, R. M\"{u}ller, G. Heger and D. Wohlleben: 
Physica C {\bf 185-189} (1991) 903.

\bibitem{Buchner93}B. B\"{u}chner, M. Cramm, M. Braden, W. Braunisch, O. Hoffels, W. Schnelle, 
R. M\"{u}ller, A. Freimuth, W. Schlabitz, G. Heger, D. I. Khomskii and D. Wohlleben: 
Europhys.\ Lett. {\bf 21} (1993) 953.

\bibitem{Kataoka86}M. Kataoka: 
J. Phys.\ C {\bf 19} (1986) 2939. 

\bibitem{Toyota88}N. Toyota, T. Kobayashi, M. Kataoka, H. F. J. Watanabe, T. Fukase, 
Y. Muto and F. Takei: 
J. Phys.\ Soc.\ Japan. {\bf 57} (1988) 3089. 

\bibitem{Kimura99}H. Kimura, K. Hirota, H. Matsushita, K. Yamada, Y. Endoh, 
S. -H. Lee, C. F. Majkrzak, R. Erwin, G. Shirane, M. Greven, Y. S. Lee, 
M. A. Kastner and R. J. Birgeneau: 
Phys.\ Rev.\ B {\bf 59} (1999) 6517.

\bibitem{Matsushita99}H. Matsushita, H. Kimura, M. Fujita, K. Yamada, K. Hirota and Y. Endoh: 
J. Phys.\ Chem.\ Solids. in press. 

\bibitem{Lee99}C. H. Lee, K. Yamada, Y. Endoh, G. Shirane, R. J. Birgeneau, 
M. A. Kastner, M. Greven. Y. J. Kim: 
To be submitted.

\bibitem{Axe73}J. D. Axe and G. Shirane: 
Phys.\ Rev.\ Lett. {\bf 30} (1973) 214.

\bibitem{Shapiro75}S. M. Shapiro, G. Shirane and J. D. Axe: 
Phys.\ Rev.\ B {\bf 12} (1975) 4899.

\bibitem{Bianconi96}A. Bianconi, N. L. Saini, A. Lanzara, M. Missori, 
T. Rossetti, H. Oyanagi, H. Yamaguchi, K. Oka and T. Ito: 
Phys.\ Rev.\ Lett. {\bf 76} (1996) 3412.

\bibitem{Billinge94}S. J. L. Billinge, G. H. Kwei and H. Takagi: 
Phys.\ Rev.\ Lett. {\bf 72} (1994) 2282.

\bibitem{Billinge96}S. J. L. Billinge and G. H. Kwei: 
J. Phys.\ Chem.\ Solids {\bf 57} (1996) 1457.

\bibitem{Wu90}Wu Ting, K. Fossheim nd T. L\ae greid:
Solid State Comm. {\bf 75} (1990) 727.

\bibitem{Wu91}Wu Ting, K. Fossheim and T. L\ae greid:
Solid State Comm. {\bf 80} (1991) 47.

\end{references}
\end{document}